\documentclass[universe,article,pdftex,moreauthors]{mdpi}  
\usepackage{bigints}
\usepackage{xcolor}
\usepackage{graphicx}
\usepackage{dcolumn}
\usepackage{bm}
\usepackage{epsfig}
\usepackage{latexsym}
\usepackage[greek,english]{babel}
\usepackage{tikz}
\usepackage{amsmath}
\usepackage{tensor}
\usepackage{amssymb}
\usepackage{caption}
\usepackage{mathrsfs}
\usepackage{braket}
\usepackage{slashed}
\usepackage{bbold}
\usepackage{float}
\usepackage{cancel}

\def\pa{\partial}

\def\calm{{\cal M}}

\def\beq{\begin{equation}}
\def\ee{\end{equation}}
\def\eeq{\end{equation}}
\def\pa{\partial}
\def\eps{\epsilon}

\def\bfig{\begin{figure}}
\def\efig{\end{figure}}
\def\bea{\begin{eqnarray}}
\def\bwt{\begin{widetext}}
\def\ewt{\end{widetext}}
\def\beann{\begin{eqnarray*}}
\def\eea{\end{eqnarray}}
\def\eeann{\end{eqnarray*}}

\def\nn{\nonumber}

\def\3p0{$^{3}P_{0}$}

\def\hs{\hspace{.5cm};\hspace{.5cm}}

\def\col#1#2#3#4 {\{ #1, #2 \}_{#3,#4}}
\def\pois#1#2 {{\{ #1, #2 \}_{q,p}=\frac{\partial #1}{\partial q_i}\,\frac{\partial #2}{\partial p_i}
-\frac{\partial #1}{\partial p_i}\,\frac{\partial #2}{\partial q_i}}}

\newcommand{\jac}[2]{\frac{\partial\left(#1\right)}{\partial\left(#2\right)}}
\firstpage{1} 
\pubvolume{1}
\issuenum{1}
\articlenumber{0}
\pubyear{2024}
\copyrightyear{2024}
\datereceived{ } 
\daterevised{ } 
\dateaccepted{ } 
\datepublished{ } 
\hreflink{https://doi.org/} 
\Title{Can Effects of a Generalized Uncertainty Principle Appear in Compact Stars?}
\TitleCitation{Title}
\Author{João Gabriel Galli Gimenez$^{1}$\orcidA{0009-0007-3254-2022},  
Dimiter Hadjimichef$^{1}$\orcidA{0000-0003-4999-7625},
Peter Otto Hess$^{2,3}$\orcidA{0000-0002-2194-7549}, 
Marcelo Netz-Marzola$^{3,5}$\orcidA{0009-0004-0384-8776},  
and 
C\'esar A. Zen Vasconcellos$^{1,4,*}$\orcidA{0000-0002-9814-5317}}
\AuthorNames{João Gabriel Galli Gimenez, Dimiter Hadjimichef, Peter O. Hess, Marcelo Netz-Marzola, C\'esar A. Zen Vasconcellos}
\AuthorCitation{Gimenez, João G. G. ; Hadjimichef, Dimiter;  Hess, Peter O.;  Netz-Marzola, Marcelo;  Zen Vasconcellos, C\'esar A.;}
\address{%
$^{1}$ \quad Instituto de F\'isica, Universidade Federal do Rio Grande do Sul (UFRGS), Porto Alegre, Brazil;cesaraugustozenvasconcellos@gmail.com\\
$^{2}$ \quad Universidad Nacional Aut\'onoma de Mexico (UNAM), M\'exico City, M\'exico;  hess@nucleares.unam.mx\\
$^{3}$ \quad  Frankfurt Institute for Advanced Studies (FIAS), Hessen, Germany \\
$^{4}$ \quad International Center for Relativistic Astrophysics Network (ICRANet), Pescara, Italy\\
$^{5}$ \quad  J.W. von Goethe Universit\"at, Frankfurt am Main, Germany
}
\corres{C\'esar A. Zen Vasconcellos: cesaraugustozenvasconcellos@gmail.com; Tel.: +5551 98357 1902}
\abstract{
In the present contribution, a preliminary analysis of the effects of the Generalized Uncertainty Principle (GUP) with a minimum length, in the context of compact stars, is performed. On basis of a deformed Poisson canonical algebra with a parametrized minimum length scale that induces deviations from conventional Quantum Mechanics, fundamental questions involving the consistence, evidences and proofs of this approach
as a possible cure for unbounded energy divergence are outlined. The incorporation of GUP effects into semiclassical 2N-dimensional systems is made by means of a time-invariant distortion transformation applied to their non-deformed counterparts.  Assuming the quantum hadrodynamics $\sigma-\omega$ approach as a toy-model, due to its simplicity and structured description of neutron stars, we perform a preliminary analysis of GUP effects with a minimum spacetime
length on these compact objects. The corresponding results for the equation of state and the mass-radius relation for neutron stars are in tune with recent observations with a maximum mass around $2.5 M_{\odot}$ and radius close to $12$ km. Our results also indicate the smallness of the noncommutative scale. 
}
\begin{document}

\section{Introduction}
The introduction of genuine gravitational effects in quantum field theory suggests an effective cutoff (minimal length) in the ultraviolet regime as a possible cure for unbounded energy divergences (ultraviolet completion).
In principle,  high-energy  probes should be sensitive to small distance
spacetime structures, revealing the quantum nature of gravity.
 The known  approaches to quantum gravity such as string
theory \cite{venez1}-\cite{gross}, loop quantum gravity \cite{lqg1}-\cite{lqg3} and quantum geometry \cite{qg} indicate, theoretically, the existence of such a minimal measurable length. In these approaches, the existence of a minimal observable length is a common
feature 
 with the estimated size of the order of the Planck length, $l_{P}\sim  10^{-33}$ cm.
A compelling way to incorporate the concept of minimum measurable length into Quantum Mechanics is to modify the algebra of the Heisenberg uncertainty principle (HUP), giving rise to the 
{\it Generalized  Uncertainty Principle} (GUP).
In the HUP framework there is essentially no restriction on the
measurement precision of the particles’ position.
In GUP, the basic phase space noncommutative algebra: $[\hat X_i,\hat P_j] \not =0$ is extended to coordinates  
$[\hat X_i,\hat X_j] \not =0$ and momenta   $[\hat P_i, \hat P_j] \not =0$.
Historically, the original idea of  extending noncommutativity to the coordinates was presented by
Heisenberg as an early attempt to  remove the infinite quantities that appear  in field theories, in a time long before the renormalization procedure was developed and gained acceptance \cite{heisenberg}.
The first analysis of a quantum theory based on noncommutative coordinates
was published by H. S. Snyder \cite{snyder}.  

In a similar approach to the quantization of a
classical phase space, a noncommutative spacetime is defined by replacing
coordinates $x^\mu$ by the Hermitian operators $\hat X^\mu$ \cite{szabo}, which in turn
obey   canonical commutation relations
\bea
[\, \hat X^\mu,\hat X^\nu\,]=i\, \,\theta^{\mu\,\nu}\,,
\label{eq2}
\eea
where   $\theta^{\mu\,\nu}$ is a constant, real-valued antisymmetric $D\times D$ matrix  
(D-dimensional spacetime) with dimensions of length squared,
that commutes with $\hat X^\mu$ and produces a coordinate uncertainty relation,
\bea
 \Delta x^\mu\,\Delta x^\nu\ \geq \frac{1}{2} \,|\theta^{\mu\,\nu}|\,.
\label{eq3}
\eea
In many occasions $\theta^{\mu\nu}$ is written as  $\theta^{\mu\nu}=\theta \,\eps^{\mu\nu}$, where $\eps^{\mu\nu}$  
 is the antisymmetric tensor and the  parameter $\theta$ defines the noncommutative scale.

A relevant question then arises:
{\it what is the size of the noncommutative scale?} Although there is no discernible definitive answer to this question, a minimum observable length in quantum field theory and quantum gravity should share the statistical properties of an expectation value. Different lower bounds have been claimed for the noncommutative scale. Conceptually, ${\theta}$ should be of the order of the Planck scale $l^2_{P}\sim 10^{-66}$ cm$^2$. However,
 when the calculation is related to the Landau problem with magnetic fields  ($\sim$12T) in the region 
of quantum Hall effect,  $\theta\sim 10^{-11}$ cm$^2$ \cite{gamboa}; 
in estimating the muon anomalous magnetic moment $\theta \sim 10^{-15}$ cm$^2$ 
\cite{wang-yang}; for   the lamb shift $\theta\sim 10^{-36}$ cm$^2$   \cite{chaichian}.

Alternatively, A. Kempf developed a deformed algebra, in which  a
modification of the canonical commutation relation was of the form \cite{kempf1}-\cite{kempf3}
\begin{equation}
    \left[\hat X,\hat P\right]=i\,\left(1+\beta\, \hat P^2\right)\ , \label{b20}
\end{equation}
where $\beta$ is a positive parameter that induces deviations from conventional Quantum Mechanics. Relation (\ref{b20}) then results in a GUP
\begin{equation}
\Delta x \Delta p \geq \frac{1}{2} \left(1 + \beta (\Delta p)^2\right)\, , 
\end{equation}
which in turn establishes the minimal length scale
\begin{equation}
    \Delta x \geq x_{min} =\sqrt{\beta}\ .
    \label{xbeta}
\end{equation}
For the $n$-dimensional case, however, the deformed Heisenberg algebra is given by the commutation relations:
\begin{align}
    &\left[\hat X_i, \hat P_j\right]=i\,\delta_{ij}\left(1+\beta \,\mathbf{\hat P}^{\,2}\right)\ ,\label{b23}\\
    &\left[\hat P_i,\hat P_j\right]=0\ ,\label{b24}\\
    &\left[\hat X_i, \hat X_j\right]=2i\beta\left(\hat X_i\,\hat P_j-\hat X_j\,\hat P_i\right) \, ,\label{b25}
\end{align}
where (\ref{b25}) defines a noncommutative geometry.
The relations given in (\ref{b23})-(\ref{b25}) do not break rotational symmetry. Indeed, the generators of rotations may still be written in terms of position and momentum operators as 
\begin{equation}
   \hat L_{ij}=\frac{1}{1+\beta \,\mathbf{p}^2}\left(\hat X_i\,\hat P_j-\hat X_j\,\hat P_i\right) \, .
\end{equation}

These algebra deformations imply profound modifications to the formalism of Quantum Mechanics. While a continuous momentum space is retained as seen from equation (\ref{b24}), the introduction of a noncommutative geometry involves the adoption of a quasi-position formalism. In this scenario, even elementary models such as the harmonic oscillator may manifest considerable complexity and notable deviations when the energy scales approach or exceed $\sqrt{\beta}$ \,\,\citep{kempf1}-\cite{kempf3}.

Among other formulations, Moayedi et al.~\cite{moayedi1,moayedi2}, introduced a deformed covariant Lorentz algebra to derive a $\beta$-modified Dirac equation, as an alternative seesaw-like mechanism for the neutrino induced by the presence of a spacetime of minimum length~\cite{dimi1}.
Marzola et al. in turn~\cite{dimi2} developed a {\it Deformed Poisson Brackets Formalism} applied to  the MIT Bag Model, introducing  
 minimal length spacetime modifications to thermodynamic quantities and in the respective equation of state.

 In the present contribution we introduce a novel analysis of GUP effects, with minimal length spacetime, in compact stars.
 It should be noted that this study comprises a still very cautious and exploratory research, where the structure of the neutron star is highly schematic.
 The main goal is to understand that {\it if} there is a GUP, with a minimal length, {\it then}
(i) the astrophysics arena represents a relevant `laboratory' to reveal these effects? (ii) what is the limiting minimal length scale?
(iii) we may identify GUP effects in neutron star data? if so, what kind of effects may be identified?

\section{The Deformed Poisson Brackets Formalism}
In what follows we present a brief summary of the formalism highlighting the classical Hamiltonian description based on Poisson brackets and the Heisenberg quantum formulation based on Dirac commutators.
\subsection{Summary of the Formalism}
\,{In Classical Mechanics,  the phase space of a given system is parametrized in terms of the canonical coordinates defined in terms of position $\mathbf{x}$ and conjugate momentum $\mathbf{p}$ $n$-tuples
\begin{equation}
\mathbf{x} = (x_1,x_2,...x_i...,x_n) \quad \mbox{and} \quad
\mathbf{p} = (p_1,p_2,...p_i...,p_n) \, . 
\end{equation}}
\,{The dynamics of the system is described by Hamilton’s equations of motion
in terms of partial derivatives of the position coordinates and the corresponding conjugate momenta: 
\begin{equation}
    \dot{x}_i = \frac{\partial H(\mathbf{x}, \mathbf{p}, t)}{\partial p_i} \quad \mbox{and} \quad
     \dot{p}_i = - \frac{\partial H(\mathbf{x}, \mathbf{p}, t)}{\partial x_i} \quad \mbox{for} \quad i= 1,2...n.
\end{equation}
where $\dot{x}_i = - \partial x_i/\partial t$ and $\dot{p}_i = \partial p_i/\partial t$ and $H(\mathbf{x}, \mathbf{p}, t)$ represents the Hamiltonian function.
The time evolution of an arbitrary function $f(\mathbf{x}, \mathbf{p}, t)$, in turn, may be expressed in terms of Hamilton's equations of motion as
\begin{eqnarray}
    \frac{d f(\mathbf{x}, \mathbf{p}, t)}{dt} & = & \sum_i^n \Biggl(
\frac{\partial f(\mathbf{x}, \mathbf{p}, t)}{\partial x_i}  \dot{x}_i
+ \frac{\partial f(\mathbf{x}, \mathbf{p}, t)}{\partial p_i} \dot{p}_i 
    \Biggr) + \frac{\partial f(\mathbf{x}, \mathbf{p}, t)}{\partial t} \nonumber \\
     & = & \sum_i^n \Biggl(
\frac{\partial f(\mathbf{x}, \mathbf{p}, t)}{\partial x_i} \frac{\partial H(\mathbf{x}, \mathbf{p}, t)}{\partial p_i}  
- \frac{\partial f(\mathbf{x}, \mathbf{p}, t)}{\partial p_i} \frac{\partial H(\mathbf{x}, \mathbf{p}, t)}{\partial x_i}   \Biggr)+ \frac{\partial f(\mathbf{x}, \mathbf{p}, t)}{\partial t}. \label{te}
\end{eqnarray}
By introducing the Poisson bracket, 
\begin{equation}
\{f(\mathbf{x}, \mathbf{p}, t), H(\mathbf{x}, \mathbf{p}, t) \} 
\equiv \sum_i^n \Biggl(
\frac{\partial f(\mathbf{x}, \mathbf{p}, t)}{\partial x_i} \frac{\partial H(\mathbf{x}, \mathbf{p}, t)}{\partial p_i}  
- \frac{\partial f(\mathbf{x}, \mathbf{p}, t)}{\partial p_i} \frac{\partial H(\mathbf{x}, \mathbf{p}, t)}{\partial x_i}   \Biggr)\, , 
\end{equation}
the eventual explicit time dependence of expression (\ref{te}) simply runs its course with the time evolution of the dynamical system
governed by the Poisson brackets binary operations: 
\begin{equation}
        \frac{d f(\mathbf{x}, \mathbf{p}, t)}{dt}   = \{f(\mathbf{x}, \mathbf{p}), H(\mathbf{x}, \mathbf{p})\} + \frac{\partial f(\mathbf{x}, \mathbf{p}, t)}{\partial t}, \label{tepb}
\end{equation}
simply considering the position and momentum dependence of the functions $f(x_i,p_i)$ and $H(x_i,p_i)$, instead of taking into account a possible explicit time dependence of these functions.
The position coordinates $x_i$ and the corresponding conjugate momenta $p_j$ obey Poisson algebra, more precisely
\begin{equation} \{x_i,p_j\}=\delta_{ij}; \quad 
   \{p_i,p_j\}=0; \quad 
    \{x_i,x_j\}=0.
\end{equation}}
\,{From a classical point of view, these relationships imply the simultaneous measurement and exact determination of the position and momentum of a particle, without uncertainties.}
\,{From this perspective, the transition from classical mechanics to quantum mechanics can be accomplished by replacing Hamilton's equations of motion by the corresponding Heisenberg equations, the classical variables ${\mathbf x}$ and ${\mathbf p}$ by the corresponding Hermitian operators in Hilbert space, $\hat{\mathbf X}$ and $\hat{\mathbf P}$, and furthermore the Poisson brackets by the Dirac commutators:
\begin{equation}
        \frac{d \hat{f}(\hat{\mathbf X}, \hat{\mathbf P}, t)}{dt}   = - \frac{i}{\hbar} [f(\hat{\mathbf X}, \hat{\mathbf P}), H(\hat{\mathbf X}, \hat{\mathbf P})] + \frac{\partial \hat{f}(\hat{\mathbf X}, \hat{\mathbf P}, t)}{\partial t}. \label{QMtepb}
\end{equation}
As a result, the Poisson algebra is replaced by the Heisenberg algebra
\begin{equation}
    \left[\hat X_i, \hat P_j\right]=i\,\delta_{ij}; \quad 
    \left[\hat P_i,\hat P_j\right]=0; \quad
    \left[\hat X_i, \hat X_j\right]=0. 
\end{equation}
which implies the Heisenberg Uncertainty Principle
\begin{equation}
    \Delta x \Delta p \geq \frac{h}{4 \pi} = \frac{\hbar}{2} . 
\end{equation}}

The Generalized Uncertainty Principle (GUP) in turn induces a deformation of the Heisenberg algebra and can be interpreted as also inducing deformations to its classical limit, which corresponds to the Poisson algebra. In what follows, in order to introduce a noncommutative algebra, we assume general deformations to the usual canonical commutation relations, such that
\begin{align}
 &\left[\hat X_i, \hat P_j\right]=i\, f_{ij}(\hat X, \hat P) \ \longrightarrow \ \{x_i,p_j\}=f_{ij}(x,p) \ ,\\
    &\left[\hat P_i,\hat P_j\right] =i\, h_{ij}(\hat X, \hat P) \ \longrightarrow \ \{p_i,p_j\}=h_{ij}(x,p) \ ,\\
    &\left[\hat X_i, \hat X_j\right]=i\, g_{ij}(\hat X, \hat P) \ \longrightarrow \ \{x_i,x_j\}=g_{ij}(x,p) \ ,
\end{align}
 where the deformation functions $f_{ij}$, $g_{ij}$ and $h_{ij}$ are restricted according to the conventional properties of commutators and brackets: obedience to the bilinearity condition, to the Jacobi identity and to the Leibniz rule provided the composition product is associative (see for instance~\cite{Amin} and references therein).
 In the particular case of the Kempf formalism, the commutation relations (\ref{b23})-(\ref{b25}), induce the following deformed Poisson brackets:
\begin{equation}
    \{x_i,p_j\}=\delta_{ij} \left(1+\beta p^2\right); \quad
    \{p_i,p_j\}=0; \quad 
    \{x_i,x_j\}=2\beta \left(p_ix_j-p_jx_i\right).
\end{equation}

\subsection{Deformation of Differential Volumes}
\,{From the perspective of Classical Statistical Mechanics, the phase space of a given physical system represents the set of all possible physical states of the system when described by a given parameterization, where each possible state corresponds uniquely to a point in the phase space represented by the coordinates $(x_i, p_i)$. Since both the position and momentum vary with time, the dynamic behavior of the system can be viewed as a continuous trajectory of points in the phase space, where the family of phase plane trajectories represents a phase portrait of the system. In Quantum Statistical Mechanics, however, the particle has no well-defined trajectory in phase space. The Heisenberg Uncertainty Principle, and therefore the Heisenberg algebra, effectively implies a discretization of the phase space in minimal volumes. A modification of the Heisenberg Uncertainty Principle (i.e., a GUP), therefore, deforms such volumes.}

Particularly in \citep{fityo2008statistical}, the phase space deformation effects are analyzed for the case of a partition function of a quantum system that is then assumed to fulfill the semiclassical limit. For a non-deformed Heisenberg algebra, we have the transition
\begin{equation}
    Z=\sum_n e^{-E_n/T} \ \longrightarrow \ \ Z=\int e^{-H(X,P)/T}\, {d^{\textit{{N}}}X\,d^{{\textit{N}}}P} \ . \label{b39}
\end{equation}

In case of a deformed algebra, it is well known that the expression for the quantum partition function will remain unaltered, since a sum over deformed volumes will have the exact same form as the sum over non-deformed ones. However, this is not the case for a classical partition function. Now we must deal with an integration over phase space that deviates from the original continuous partition function in (\ref{b39}) having as the Jacobian factor
\begin{equation}
    J=\frac{\partial(x,p)}{\partial(X,P)} \,.
\end{equation}
This Jacobian factor distorts the phase space, relating the canonical variables of the non-deformed algebra (which we called $X$ and $P$) to the ones of the deformed algebra $x$ and $p$. Namely, we have 
\begin{equation}
    Z=\sum_n e^{-E_n/T} \ \longrightarrow \ \ Z=\int e^{-H(x,p)/T}\, \frac{d^{\textit{N}}x\,d^{\textit{N}}p}{J} \, .
\end{equation}

An important canonical invariant is the magnitude of a volume element in phase space. 
A canonical procedure   transforms the $2N$-dimensional phase space with coordinates $\eta_i$
to another phase space with coordinates $\zeta_i$ \cite{goldstein}. In this sense, the Jacobian of the transformation between non-deformed and deformed algebras can be written purely as combinations of deformed Poisson brackets, and, in the particular case of the Kempf deformed algebra of a $2N$-dimensional phase space, we have \citep{goldstein,fityo2008statistical,chang2002effect}
\bea
    J=\prod^{N}_{i=1}\,\{x_i,p_i\}=\left(1+\beta p^2\right)^N \ .
\label{jtrans1}
\eea
This is an important result that gives us the possibility to calculate (for general deformations) the continuous function without introducing canonically conjugated auxiliary variables. 

In other words, we may effectively incorporate the effects of a GUP into semiclassical $2N$-dimensional systems by applying the following transformation to their non-deformed counterparts:
\begin{equation}
    d^{\textit{{N}}}x\,d^{\textit{{N}}}p \ \longrightarrow \ \frac{d^{\textit{\tiny{N}}}x\,d^{\textit{\tiny{N}}}p}{(1+\beta p^2)^N} \, ,
    \label{b43}
\end{equation}
which represents a distortion in the differential volumes of phase space and can be shown to be invariant under time evolution from  the Liouville theorem \citep{chang2002effect}. The effective formalism described here has been derived equivalently as a deformation of the Planck constant $h$ in~\citep{rama2001some}.

 \section{The QHD-I Model}
  
\,{Quantum hadrodynamics $\sigma-\omega$ model (QHD-I)~\cite{qhd1}-\cite{qhd5}, as a relativistic quantum field theory for baryons and mesons, has been
widely applied to studying various nuclear phenomena including the hadron-hadron interaction, the hadron-nucleus
scattering, the bulk and single-particle properties of nuclei, among others. It is commonly recognized that although
the quantum chromodynamics is a fundamental theory for strong interaction, the QHD, as an effective field theory
formulated in terms of hadronic degrees of freedom, provides a simple and reliable approach to produce the nuclear
observables that are insensitive to the short-range dynamics. There are various QHD models, renormalizable and
nonrenormalizable, which were tested in the past to reproduce the empirical nuclear properties and the experimental
data. In particular, the $\sigma-\omega$  model proposed by Walecka \cite{qhd1,qhd2} contains nucleons with the parametrized mass denoted as $M$ and Lorentz isoscalar-scalar mesons $\sigma$ and isoscalar-vector mesons $\omega$.}
The nonrelativistic approximations leads to a nucleon-nucleon interaction potential which behaves as short-range repulsion
and medium-range attraction.  The model considers that the
central effective potential for the nucleon-nucleon interaction is
given by 
\bea
V(r)=\frac{g_\omega^2}{4 \pi}\frac{e^{-m_\omega r}}{r} - \frac{g_{\sigma}^2}{4 \pi}
\frac{e^{-m_{\sigma} r}}{r},
\eea
where $r$ defines the relative
distance between two nucleons, the two constants $g_\sigma$ and $g_\omega$
are adjusted to reproduce the nucleon-nucleon interaction and the
meson masses are respectively $m_\sigma=550$ MeV and $m_\omega=783$ MeV.

The  $\sigma-\omega$ model can be summarized in a nutshell, starting with the   model lagrangian defined as 
 \begin{eqnarray}
  \mathcal{L}& =& \bar{\psi}(i\,\gamma^\mu \,D_\mu - \calm)\psi + \frac{1}{2}(\partial_\mu \sigma \partial^\mu \sigma - m_\sigma^2 \sigma^2) 
      -\frac{1}{4}F_{\mu\nu}F^{\mu\nu} + \frac{1}{2}m_\omega^2\, \omega_\mu \omega^\mu  
\label{ws1}
 \end{eqnarray}
where $\psi$ denotes the nucleon wave-function and
\bea
D_\mu=\pa_\mu +i\,g_\omega\,\omega_\mu  \hs    \calm = M - g_{\sigma}\, \sigma \hs   
F_{\mu \nu} = \partial_{\mu} \omega_{\nu} - \partial_{\nu} \omega_{\mu}
\label{ws2}
\eea
obtaining the following equations (with $\square=\partial_{\mu} \partial^{\mu}$)
\bea
  (\square + m_{\sigma}^2) \sigma = g_{\sigma}\,\bar{\psi}\,  \psi \hs
     \partial_{\mu} F^{\mu \nu} + m_{\omega}^2 \,\omega^{\nu} = g_{\omega} \,\bar{\psi} \gamma^{\nu} \psi
     \hs
     \left(\,i\, \gamma^{\mu} D_\mu - \calm \,\right) \psi = 0
\label{ws3}
\eea
\,{The parameters $M$, $g_{\sigma}$, $g_{\omega}$, $m_{\sigma}$, and $m_{\omega}$ are phenomenological constants that may be determined (in principle) from experimental observables. }

Assuming static and uniform nuclear matter, in its ground state and the mean field approximation (MFA), nucleons can be seen as under the action of an average nuclear interaction and operating only with the space and time c-number values of the meson fields. The higher the baryonic density, the better the validity of this approximation, since, at high densities, the fluctuations of the meson fields are negligible when compared with the amplitudes of the nucleon fields, which allows them to be replaced by their expected mean values. The mean field approximation is then used~\cite{qhd2}, which allows the deduction of a semi-analytical solution and which consists of using the average classical values of the meson fields:
\bea
 \sigma \to \langle \sigma \rangle \equiv \sigma_0\hs
  \omega^{\mu} \rightarrow \langle \omega^{\mu} \rangle \equiv \delta_{\mu 0} \,\omega_0\hs
  \calm \to  M^{\ast} = M - g_{\sigma} \,\sigma_0
  \langle F_{\mu \nu} \rangle = 0\,.
\label{ws4}
\eea
In this expression, $M^{\ast}$ represents the nucleon effective mass.
Substituting (\ref{ws4}) in (\ref{ws3}) one obtains 
\bea
\sigma_0 = \frac{g_{\sigma}}{m_{\sigma}^2} \,\rho_s \hs  \omega_0 = \frac{g_{\omega}}{m_{\omega}^2} \,\rho_B\,,
\label{ws5}
\eea
\,{with $\rho_s \equiv \langle \bar{\psi} \psi \rangle$, which represents the scalar density, and $ \rho_B = \langle \psi^{\dagger} \psi \rangle$, which denotes the baryon density.} The equation of state (EoS) is calculated from
the mean field energy-momentum tensor $\langle T_{\mu\nu} \rangle_{\rm  MFA}$ 
and in summary can be written as a function of the Fermi momentum $k_F$:
\bea
     p&=&-\frac{1}{2}m_\sigma^2 \sigma_0^2+\frac{1}{2}m_\omega^2\omega_0^2+\frac{1}{3} \frac{\gamma}{(2 \pi)^3} \int_{0}^{k_F} \frac{k^2}{\sqrt{M^{*2} + k^2}} \, d^3k\,,
\nn\\
 \epsilon &=& \frac{1}{2} m_\sigma^2 \sigma_0^2 - \frac{1}{2} m_\omega^2 \omega_0^2 + \frac{\gamma}{(2 \pi)^3} \int_{0}^{k_F} (\sqrt{k^2 + M^{*2}}+g_\omega \omega_0) \, d^3k\,,
 \label{eos-sw0}
\eea
where after integration results in 
\bea
 p&=&- \frac{1}{2}\frac{m_\sigma^2}{g_\sigma^2}(M - M^{*})^2 + \frac{\gamma^2}{72 \pi^4} \frac{g_\omega^2}{m_\omega^2}k_F^6  \nonumber \\
    &&+ \frac{\gamma}{6\pi^2}\left[\left( \frac{1}{4}k_F^3 - \frac{3}{8}M^{*2} k_F \right) \sqrt{M^{*2} + k_F^2} + \frac{3}{8}M^{*4} \ln \frac{k_F + \sqrt{M^{*2} + k_F^2}}{M^{*}}\right]\,,
     \label{eos-sw1}
\eea
and
\bea
  \epsilon &=& \frac{1}{2}\frac{m_\sigma^2}{g_\sigma^2}(M - M^{*})^2 + \frac{\gamma^2}{72 \pi^4} \frac{g_\omega^2}{m_\omega^2}k_F^6 
   \nonumber \\
   && + \frac{\gamma}{2\pi^2} \left[ \left( \frac{1}{8} M^{*2} k_F + \frac{1}{4} k_F^3 \right) \sqrt{M^{*2} + k_F^2} - \frac{1}{8} M^{*4} \ln \frac{k_F + \sqrt{M^{*2} + k_F^2}}{M^{*}} \right]\,.
   \label{eos-sw2}
\eea
The baryon density and the \,{nucleon} effective mass are also written in terms of $k_F$
\bea
\rho_B &=& \frac{\gamma}{(2\pi)^3} \int_{0}^{k_F} d^3k= \frac{\gamma k_F^3}{6\pi^2}
\nn\\
 M^{*}&=&M - \frac{g_\sigma^2}{m_\sigma^2} \frac{\gamma M^{*}}{2\pi^2}\left[ \frac{1}{2} k_F \sqrt{M^{*2} + k_F^2} - \frac{1}{2}M^{*2} \ln \left( \frac{k_F + \sqrt{M^{*2} + k_F^2}}{M^{*}} \right) \right]\,.
\eea
At this point it is possible to effectively incorporate the effects of a GUP into the semiclassical system by applying the transformation
(\ref{b43}) to the non-deformed phase space:
\begin{equation}
    d^{\textit{{3}}}x\,d^{\textit{{3}}}p \ \longrightarrow \ \frac{d^{\textit{\tiny{3}}}x\,d^{\textit{\tiny{3}}}p}{(1+\beta p^2)^3}
    \approx (1-3\,\beta p^2) \,    d^{\textit{{3}}}x\,d^{\textit{{3}}}p ,
    \label{jtrans2}
\end{equation}
where the approximation is justified\footnote{\,{Concerning the approximation $\beta\,p^2<<1$, since no large values of $k_f$ should be taken into account  in our calculations, - according to the range of values adopted for conventional formulations of equations of state of neutron stars -, our corresponding evaluation is based on the following  preliminary estimate: 
    the maximum value of $k_f$ is $5 \times $fm$^{-1}$ which leads to the maximum value of $\beta$ equal to $4 \times 10^{-2}$fm$^2$.
    In case we assume $\beta = 5 \times 10^{-2}$ fm$^2$ then the corresponding value of $k_f$ goes up to $4$ fm.}} considering  $\beta\,p^2\ll 1$. 
The EoS in Eq. (\ref{eos-sw0}) modification results in the following expressions\footnote{\,{In quantum field theory, since we are using a finite normalization volume $V$, we should be summing over a group of allowed wave vectors $\mathbf{k}$, for large volume (see~\cite{Mandl})
$$
\Biggl(1/ \int d^3 x  \Biggr) \times \sum_{\mathbf{k}} \rightarrow \frac{1}{(2 \pi)^3} \int d^3 \mathbf{k}
$$
with $\mathbf{p} = \hbar \mathbf{k}$ and $\hbar = 1$. Accordingly, the normalization volume $V$ should drop out of all physically significant quantities. In these equations, the angular part in $d^3 k$ was integrated leaving only the `radial' part in momentum space according to $ d^3 k = 4 \pi k^2 dk.$}}
\bea
     p_{\beta}&=&-\frac{1}{2}m_\sigma^2 \sigma_0^2
     +\frac{1}{2}m_\omega^2\omega_0^2+ \frac{\gamma}{6\pi^2} \int_{0}^{k_F} \frac{k^4}{\sqrt{M^{*2} + k^2}}(1-3\beta k^2) dk
\nn\\
    \epsilon_{\beta} &=& \frac{1}{2} m_\sigma^2 \sigma_0^2 
    - \frac{1}{2} m_\omega^2 \omega_0^2 + \frac{\gamma}{2 \pi^2} \int_{0}^{k_F} k^2\sqrt{k^2 + M^{*2}}(1-3\beta k^2)dk + g_\omega \,
    \omega_0\, \rho_{B\beta}\,.
 \label{eos-sw3}
\eea
After integration, these expressions result in
 \bea
     p_{\beta}& &= - \frac{1}{2}\frac{m_\sigma^2}{g_\sigma^2}(M - M^{*})^2 + \frac{\gamma^2}{72 \pi^4} \frac{g_\omega^2}{m_\omega^2}k_F^6 -  \frac{\gamma^2 \beta}{20 \pi^4} \frac{g_\omega^2}{m_\omega^2}k_F^8  \nonumber \\
     &&+ \frac{\gamma}{6\pi^2}\left[\left( \frac{1}{4}k_F^3 - \frac{3}{8}M^{*2} k_F \right) \sqrt{M^{*2} + k_F^2} + \frac{3}{8}M^{*4} \ln{\left(\frac{k_F+\sqrt{M^{*2}+k_F^2}}{M^{*}}\right) } \right] \nonumber \\
     &&- \frac{\gamma \beta}{2 \pi^2}\left[\left(\frac{1}{6}k_F^5-\frac{5}{24}M^{*2}k_F^3+\frac{5}{16}M^{*4} k_F\right)\sqrt{M^{*2}+k_F^2}
     \right.\nn\\
     &&\left.     -\frac{5}{16}M^{*6} \ln{\left(\frac{k_F+\sqrt{M^{*2}+k_F^2}}{M^{*}}\right) } \right]\,,
\label{eos-sw4}
\eea
and
\bea
    \epsilon_{\beta} &=& \frac{1}{2}\frac{m_\sigma^2}{g_\sigma^2}(M - M^{*})^2 + \frac{\gamma^2}{72 \pi^4} \frac{g_\omega^2}{m_\omega^2}k_F^6  -  \frac{\gamma^2 \beta}{20 \pi^4} \frac{g_\omega^2}{m_\omega^2}k_F^8 \nonumber \\ 
    &&+ \frac{\gamma}{2\pi^2} \left[ \left( \frac{1}{8} M^{*2} k_F + \frac{1}{4} k_F^3 \right) \sqrt{M^{*2} + k_F^2} - \frac{1}{8} M^{*4} \ln{\left(\frac{k_F+\sqrt{M^{*2}+k_F^2}}{M^{*}}\right) } \right] \nonumber \\
    &&- \frac{3\gamma \beta}{2 \pi^2}\left[(\frac{1}{6}k_F^3 - \frac{M^{*2} k_F}{8})(M^{*2}+k_F^2)^{\frac{3}{2}}+\frac{M^{*4}}{16}k_F \sqrt{M^{*2}+k_F^2}
     \right.\nn\\
     &&\left.
    +\frac{M^{*6}}{16} \ln{\left(\frac{k_F+\sqrt{M^{*2}+k_F^2}}{M^{*}}\right)} \right]\,.
 \label{eos-sw5}
\eea
  The new baryon density and the nucleon effective mass may be also written as      
\bea
\rho_{B\beta}&=& \frac{\gamma k_F^3}{6\pi^2} - \frac{3}{10}\frac{\beta \gamma}{\pi^2}k_F^5
\nn\\
        M^{*}&=& M -  \frac{g_\sigma^2}{m_\sigma^2} \frac{\gamma M^{*}}{4\pi^2} \left( k_F \sqrt{M^{*2} + k_F^2} - M^{*2} \ln \left( \frac{k_F + \sqrt{M^{*2} + k_F^2}}{M^{*}} \right) \right) \, , \nonumber \\
     &&   + \frac{g_\sigma^2}{m_\sigma^2} \frac{\gamma 3 M^{*} \beta}{2\pi^2} \left[\left( \frac{1}{4}k_F^3 - \frac{3}{8}M^{*2} k_F \right) \sqrt{M^{*2} + k_F^2} 
      \right.\nn\\
     &&\left.
     + \frac{3}{8}M^{*4}  \ln \left( \frac{k_F + \sqrt{M^{*2} + k_F^2}}{M^{*}} \right) \right] \,.
\label{eos-sw6}
\eea
 
\section{Results and Discussion}

\begin{figure}[tbph]
\centering
\includegraphics[width=12 cm]{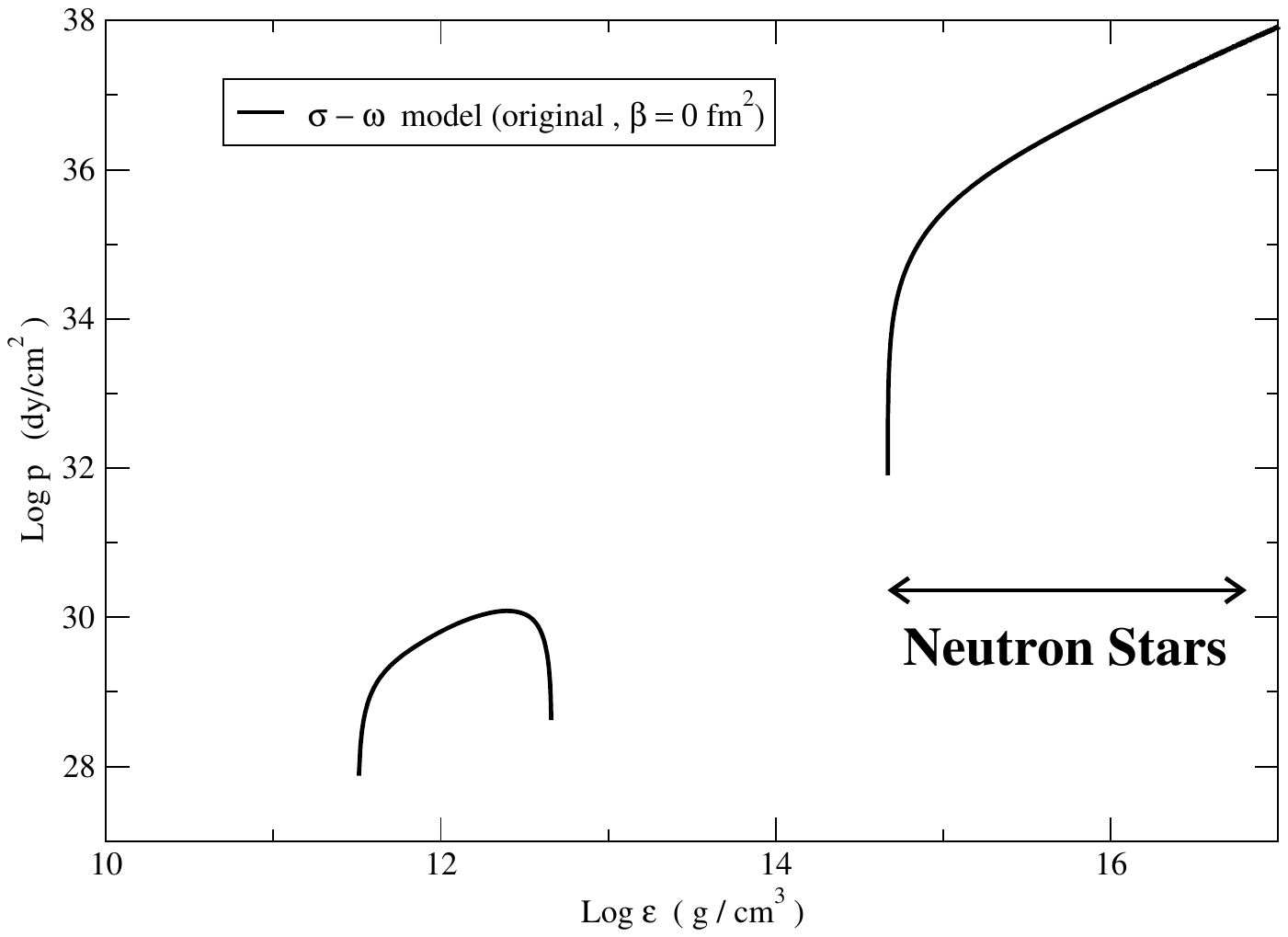}
\caption{\,{Phase transition in nuclear matter similar to the liquid-gas transition for $\gamma = 4$. $\gamma$ indicates the spin-isospin degeneracy factor, equal to $4$ for symmetric matter ($N = Z$) and $2$ for pure neutron matter ($Z = 0$). The plot shows the pressure $p$ as a function of the energy density $\varepsilon$ in logarithmic scale. The curve corresponds to the commutative $\sigma-\omega$ model, where $\beta=0$, representing the original case. The region marked "Neutron Stars" indicates the typical range of energy density relevant for neutron star matter.}
\label{fig1}}
\end{figure}

\begin{figure}[tbph]
\centering
\includegraphics[width=12 cm]{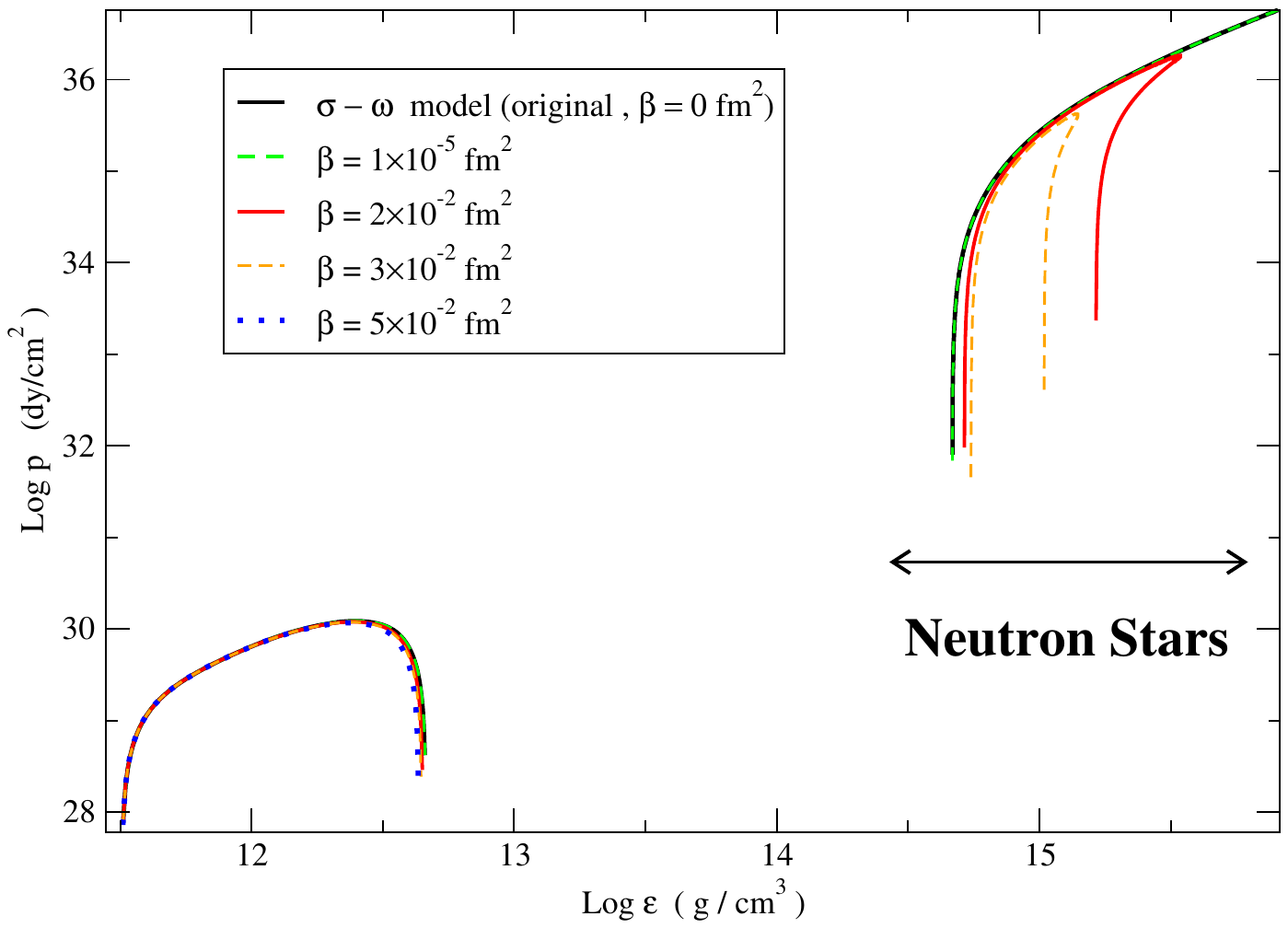}
\caption{\,{Noncommutative phase transition for nuclear matter in a logarithmic pressure vs. energy density plot. The curves correspond to different values of the noncommutativity parameter $\beta$, as indicated in the legend. The figure shows a limit of $\beta = 5 \times 10^{-2} \, \mathrm{fm}^2$, beyond which the phase transition ceases to exist.}
\label{fig2}}
\end{figure}

\begin{figure}[tbph]
\centering
\includegraphics[width=12 cm]{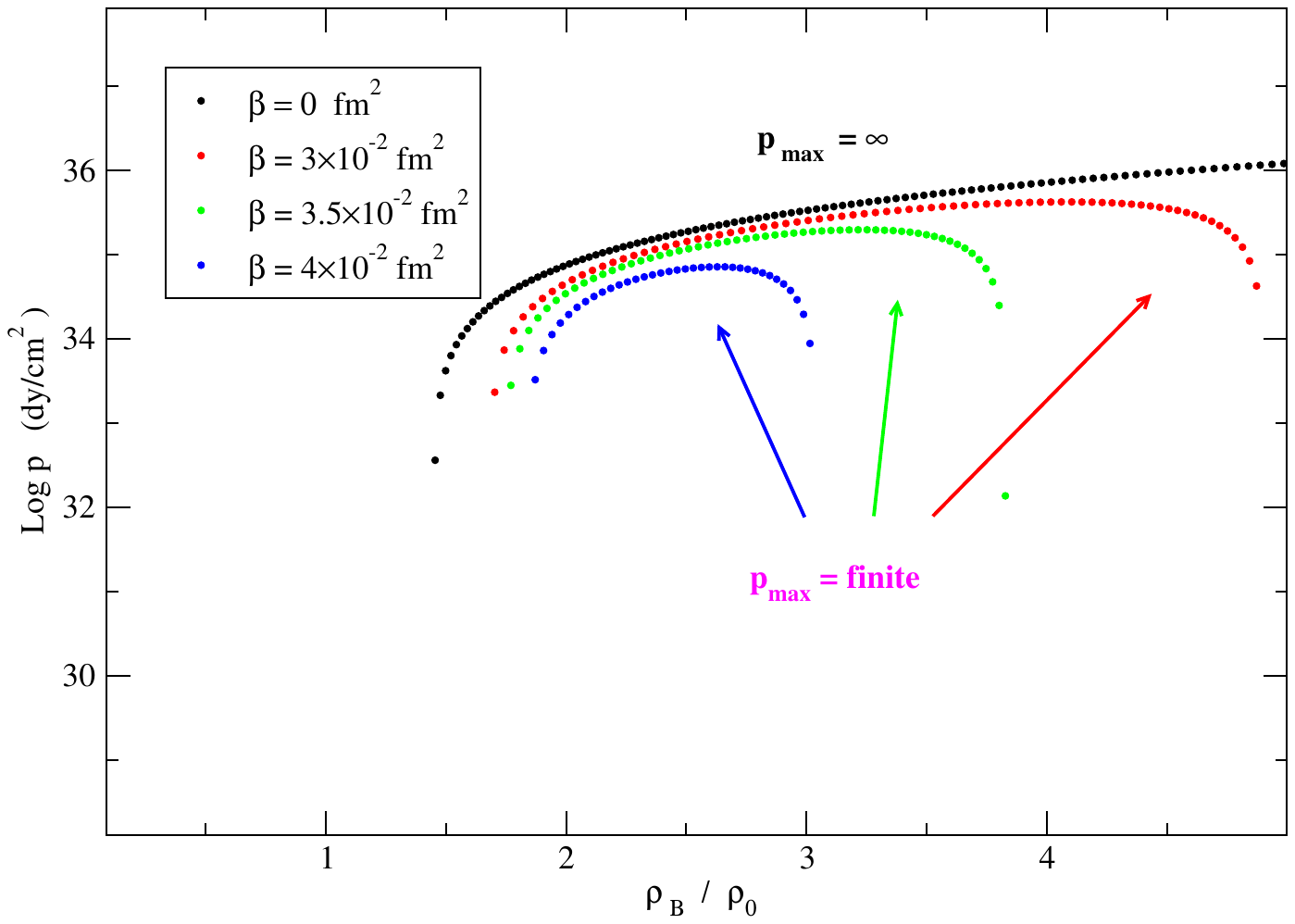}
\caption{\,{Maximum values of pressure as a function of the baryon density $\rho_B / \rho_0$ in a noncommutative geometry model for nuclear matter. The curves represent different values of the noncommutativity parameter $\beta$. For finite $\beta$, the pressure reaches a maximum value, contrasting with the original commutative case where $p_{\text{max}} = \infty$. The colored arrows highlight the respective curves for increasing $\beta$.}
\label{fig3}}
\end{figure}

\begin{figure}[tbph]
\centering
\includegraphics[width=12 cm]{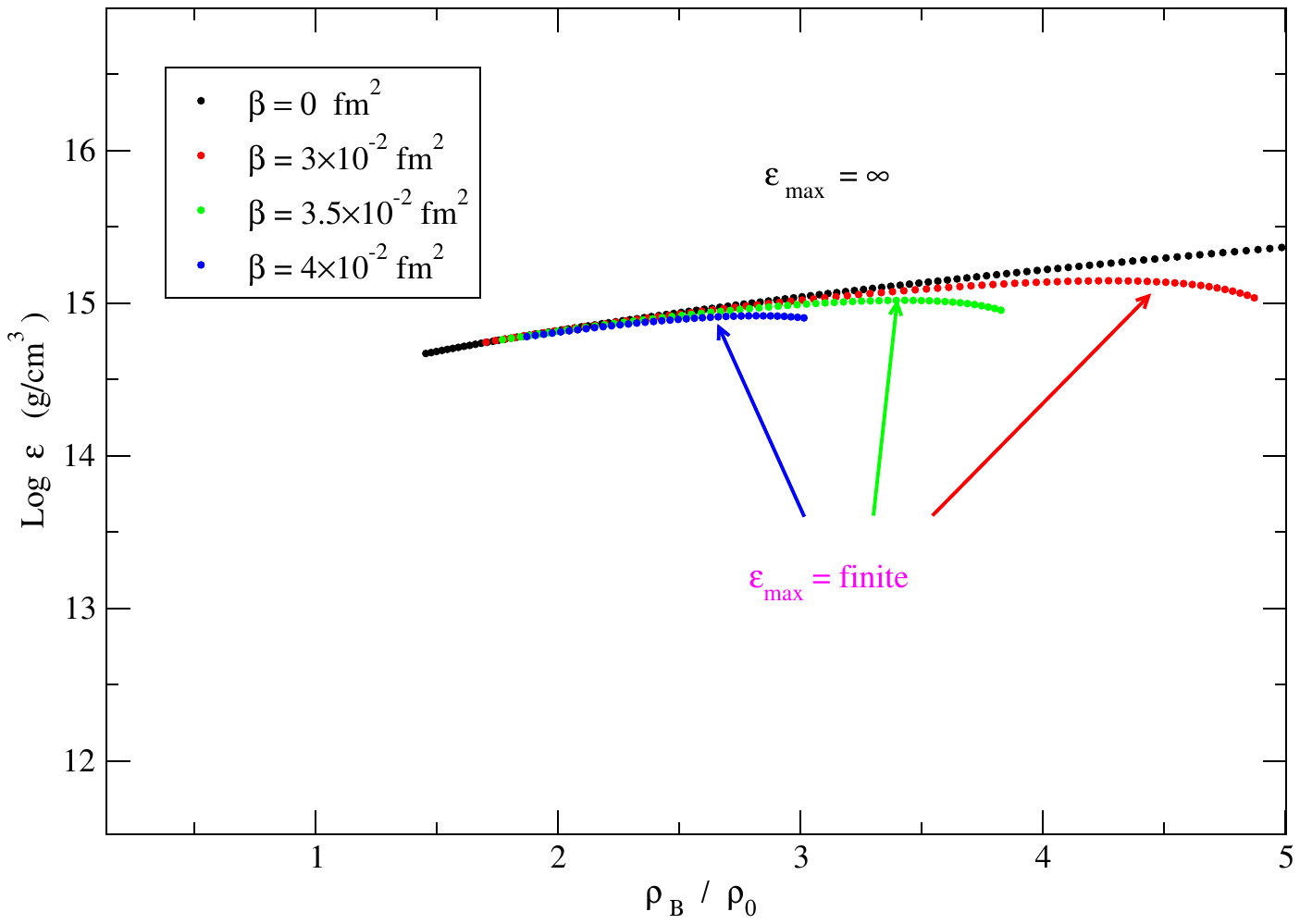}
\caption{\,{Maximum values of energy density as a function of the baryon density $\rho_B / \rho_0$ in a noncommutative geometry model for nuclear matter. The curves correspond to different values of the noncommutativity parameter $\beta$. For finite $\beta$, the energy density reaches a maximum value, contrasting with the original commutative case where $\varepsilon_{\text{max}} = \infty$. The colored arrows highlight the behavior of the curves for increasing $\beta$.}
\label{fig4}}
\end{figure}

\begin{figure}[tbph]
\centering
\includegraphics[width=9.5 cm]{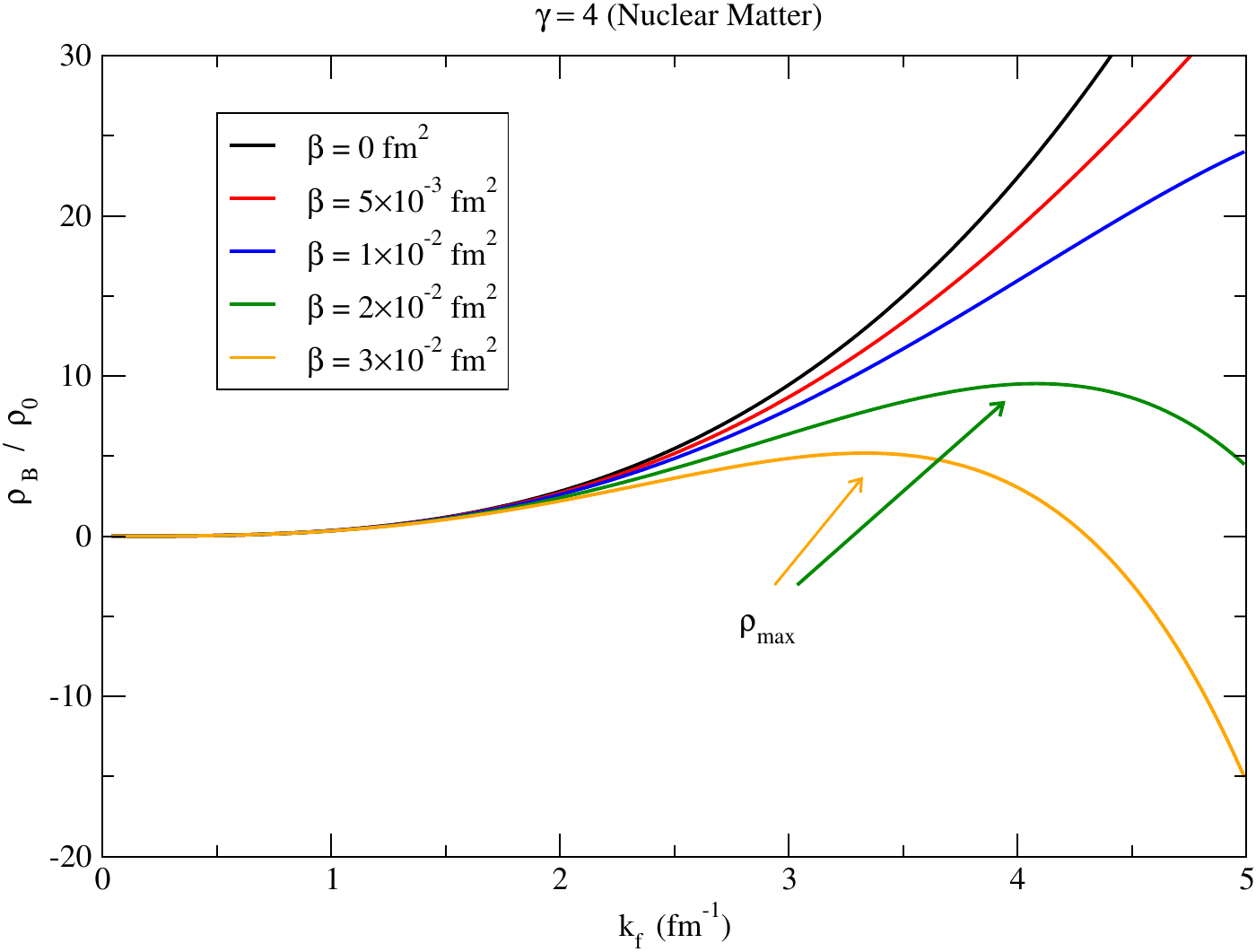}
\caption{\,{Baryon density $\rho_B / \rho_0$ as a function of the Fermi momentum $k_F$ in a noncommutative geometry model for nuclear matter.  The curves represent different values of the noncommutativity parameter $\beta$. For larger $\beta$, the baryon density reaches a maximum value $\rho_{\mathrm{max}}$, as indicated by the green and orange curves, before decreasing.}
\label{fig5}}
\end{figure}

\begin{figure}[tbph]
\centering
\includegraphics[width=9.5 cm]{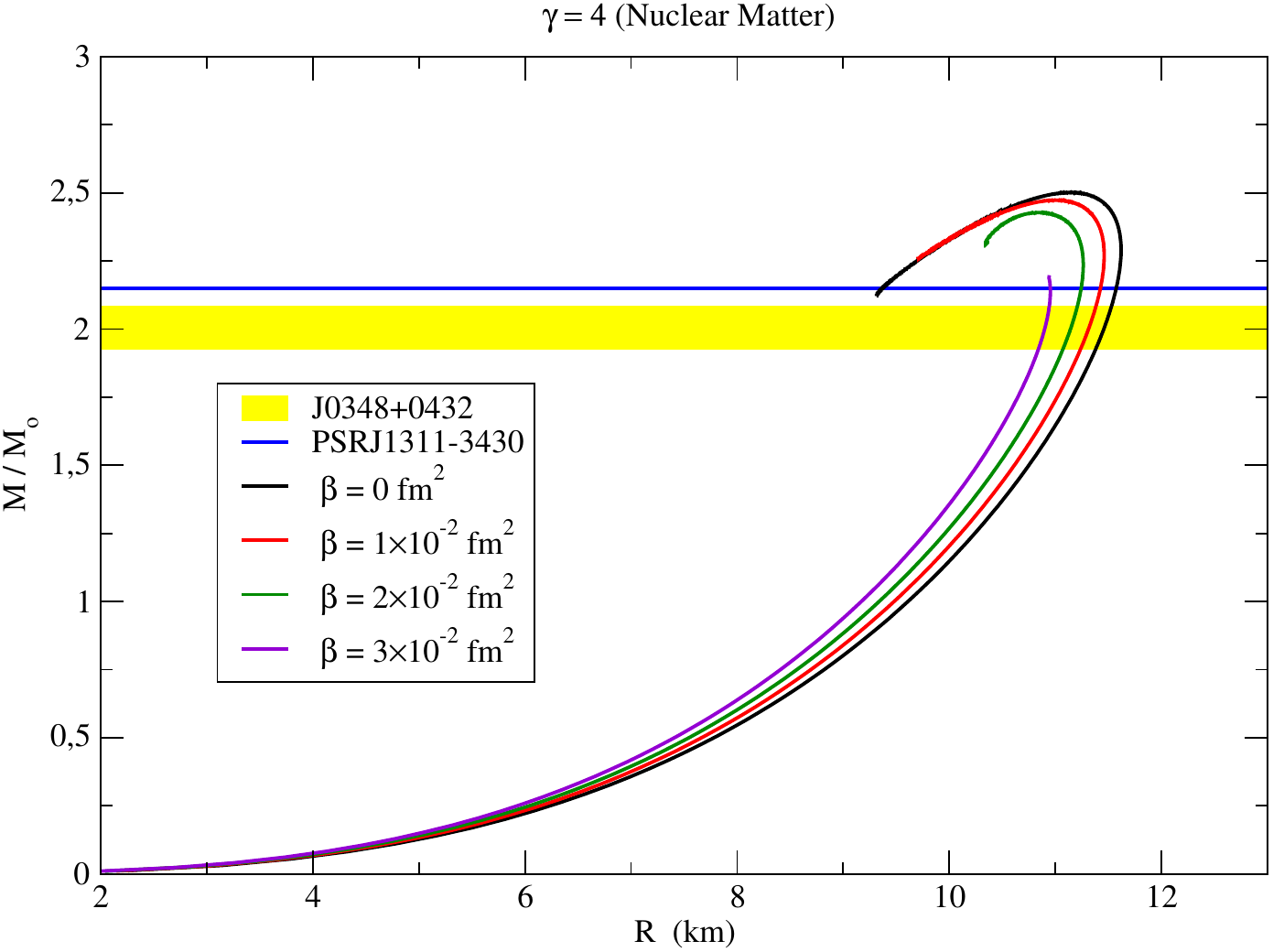}
\caption{\,{Mass–radius diagram for compact stars obtained for various values of the noncommutativity parameter $\beta$. 
    The curves correspond to different $\beta$. The horizontal yellow band represents the mass constraint from the pulsar J0348+0432, while the blue line corresponds to the mass of PSR J1311-3430. The figure demonstrates that increasing $\beta$ reduces the maximum mass of the neutron stars.}
\label{fig6}}
\end{figure}

In this section we shall explore the novel effects generated by the noncommutative spacetime GUT deformation applied to 
the Walecka's $\sigma-\omega$ formulation for neutron stars, assumed as a kind of toy-model in view of its formal simplicity.
\,{It is well known that this model predicts a phase transition similar to the liquid-gas transition of the van der Waals equation of state. Moreover, the coexisting pressure and binodal density properties of the two phases are deduced by means of a Maxwell construction applied to the equation of state of nuclear matter, obtained by employing the principle of least action to the QHD-I Lagrangian density.  At high densities, the system approaches the causal limit $p = \eps$, representing the
"stiffest" possible equation of state, as can be seen in figure~\ref{fig1}. Despite its formal simplicity, a relevant aspect to be highlighted in this theoretical approach, even if one considers more complete models from the point of view of inserting in the Lagrangian density, for instance, the fundamental meson octet and baryon decuplet, the dynamics generated by scalar and vector mesons will remain significantly present. This is because the neutral scalar and vector components coupled to the nucleons are the most relevant ingredients for describing nuclear properties in bulk, which is our main concern here.}

\,{The results in turn corresponding to the insertion of the GUT deformation into the QHD-I model are presented in figures \ref{fig2}-\ref{fig6}.}

\,{The GUP establishes that the noncommutative spacetime be dependent on a minimal length parameter $\sqrt{\beta}$ as previously highlighted in equation (\ref{xbeta}), whose size ordering is unknown. Therefore, the strategy here has been to consider it as a free parameter and identify noticeable changes in related observables at different scales, chosen consistently from the Planck scale, which represents the lower limits of our current understanding of quantum mechanics, to the typical dimensions of neutron stars.}

\,{In this domain, the smallest physical scale corresponds to the Planck domain with $\sqrt{\beta}\sim  l_{Planck}$. This scale is far below typical neutron stars observable range of values which are of the inverse of Fermi-momentum order, $k_F^{-1}$. In the sequence, to make contact with reference calculations~\cite{gamboa,wang-yang,chaichian} and  consider larger effective noncommutative scales, we limit the $\beta$ values to the range $[1\times 10^{-5}$ fm$^2-5\times 10^{-2}$ fm$^2]$.}
  
\,{After solving equations (\ref{eos-sw4}) and (\ref{eos-sw5}) for the noncommutative EoS, together with the equation for the nucleon effective mass (\ref{eos-sw6}), the corresponding results are shown in figure~\ref{fig2} for various values of $\beta$ exhibiting the behavior of the noncommutative phase transition for nuclear matter. The figure shows that the phase transition for nuclear matter is present, as in the usual model (figure~\ref{fig1}), but with a new feature: the pressure has a maximum value and is "squeezed down" with increasing $\beta$. There is a limit of $\beta=5\times 10^{-2}$ fm$^2$ after which the phase transition ceases to exist. The same effects appear in figures ~\ref{fig3} and \ref{fig4} for the pressure and energy density as a function of the baryon density, where the limiting maximum values are evident. In figure \ref{fig5}, the baryon density is plotted as a function of $k_F$, again exhibiting a squeezing down of maximum values, now for $\rho_B$. Here an interesting interpretation can be put forward: when probing short distances, which correspond to larger $k_F$, in a spacetime that has a large minimum value $\beta$, the system becomes dilute and the density drops with $k_F$.}

\,{The consistently observed behavior of decreasing the maximum values of pressure, energy density, and baryon density as far as the $\beta$ value increases can be understood through an analysis of the effects of the minimum scale into the phase space volume of the solutions 
of the equations addressed in a comparison and analogy with the corresponding solutions associated with the motion of a particle in quantum mechanics.
In quantum mechanics, as is well known, the uncertainty principle states that it is not possible to measure the position and momentum of a particle with absolute precision. A state of motion can only be given with this indefinitiness and corresponds in phase space to an elementary cell volume of size $(2 \pi \hbar)^3$. The number of quantum states available to a particle will therefore be finite and corresponds to the total volume of the phase space divided by the size of the elementary cell
\begin{equation}
    {\cal N} = \frac{1}{(2 \pi \hbar)^3} \int dx \, dy \, dz \, dp_x \, dp_y \, dp_z. 
\end{equation}}
\,{In the present case the number of quantum states available consistent with a coherent description of the properties of a neutron star obey, due to the presence of a minimum scale, a relation of the type
\begin{equation}
    {\cal N}_{Neutron \, Star} = \frac{1}{(2 \pi \hbar)^3} V^{\beta}_{ps} = \frac{1}{(2 \pi \hbar)^3}  \int \frac{d^{\textit{\tiny{N}}}x\,d^{\textit{\tiny{N}}}p}{(1+\beta p^2)^N} \, .
\end{equation}}
\,{The consequences of this transformation of the phase space volume $V^{\beta}_{ps}$ due to the presence of the minimum scale parameter, $\beta$ are evident in that $V^{\beta}_{ps}$ decreases as $\beta$ increases, with exactly the contrary occurring in the opposite case. This means that in the case where the volume of the phase space increases, corresponding to the decrease of $\beta$, more solutions consistent with a coherent description of the properties of a neutron star find more space for their realization. And the opposite occurs when $\beta$ increases. Therefore, as a consequence, the maximum values for the quantities previously plotted become dependent on the size of the phase space of the solutions which in turn depend on the parameter $\beta$, with the largest values of their amplitudes corresponding to the smallest values of the minimum scale parameter, and the contrary occurring in the opposite case.}

\,{Similarly, the same occurs in the curves corresponding to the maximum mass of neutron star families, with their values squeezing down with increasing $\beta$, as we will see below.
In the usual approach, the essential nuclear physics ingredients for astrophysical calculations
are appropriate equations of state (EoS). After the EoSs are chosen, they enter as input to
the Tolman–Oppenheimer–Volkoff equations (TOV), which in turn give as output
some macroscopic stellar properties: radii, masses, and central energy densities.
This may raise a philosophical question of {\it at what level should the noncommutative effects  be introduced?}
If one follows the idea of first principles, then Einstein's field equations should be modified to bring this information,
originated at a level of quantum gravity. This is a long step, still far beyond the current knowledge. 
The procedure we follow will be more conservative and consider that the classical field equations remain valid, only the EoS will be modified by the noncommutative spacetime. The TOV equations then become:
\bea
\frac{dp_\beta}{dr}&=&-\frac{\left[p_\beta(r)+\eps_\beta(r) \right]\, [M_\beta(r)+4\,\pi r^3\,p_\beta(r)]}{r[r-2M_\beta(r)]}
\nn\\
\frac{dM_\beta}{dr}&=&4\,\pi  \,r^2 \,\eps_\beta(r )\,.
\label{tov}
\eea}
\,{Solving the TOV equations for the EoS (\ref{eos-sw4}) and (\ref{eos-sw5}),
 together with the effective mass expression (\ref{eos-sw6}) results in the Mass-Radius diagram of
 figure~\ref{fig6}. The noncommutative effect that was seen in the EoS figures  is again present, limiting
 the neutron star's maximum mass.
In the figure, for comparison, the experimental values of PSR J0348+0432,
a pulsar-white dwarf binary system in the constellation Taurus,  with mass of 2.01 $\pm$ 0.04 M$_\odot$
and PSR J1311–3430, a pulsar with a spin period of 2.5 milliseconds, with 
mass of 2.15 M$_\odot$.}
\unskip
\unskip
\unskip
\unskip
\unskip
\,{One aspect to be highlighted concerns the complementation of the curves that describe the behavior of stellar masses as a function of their radii. As $\beta$ increases, the curves corresponding to the relation $M \times R$ become less and less complemented, arriving even to the limit of not reaching, in the case where $\beta = 3 \times 10^{-2}$fm$^2$, their maximum mass value. This can be understood to the extent that increasing the value of the parameter $\beta$ will imply a decrease in both the radii and the stellar masses, for the reasons identified previously, this occurring mainly in the region of the `tails' of the amplitudes corresponding to the relation $M \times R$. In this region, the significant decrease in pressure contributes to a decrease in the complementation of the corresponding curves and, ultimately, to preventing the maximum mass from being obtained.}

\section{Conclusions}

\,{As stated before, this study was very cautious and exploratory applied to a schematically structured neutron star model.
The model adopted for a neutron star, as emphasized previously, due to its formal simplicity and structured characteristics was chosen in order to perform a preliminary analysis of the GUP effects with a minimum length  for the description of the spacetime noncommutativity effects.  }

\,{The adopted model is well known as Quantum Hadrodynamics (QHD-I), a relativistic quantum field theory based on a local Lagrangian density with couplings between nucleons and mesons of the attractive scalar-isoscalar $\sigma$ and repulsive vector-isoscalar $\omega$ types playing the role of the relevant mean-field effective mesonic degrees of freedom. This model, despite its formal simplicity, provides a consistent theoretical framework to describe global static properties of many-body systems under the action of the strong interaction and in extreme conditions of pressure and density such as those found in neutron stars and pulsars. The limitations of the model, among others, are reflected in the description of some of the static properties of nuclear matter, such as the effective mass of the nucleon and the compressibility of symmetric nuclear matter. In this sense, future theoretical calculations indicate the need to propose a model for neutron stars that contemplates a phenomenological Lagrangian formalism with nonlinear scalar, vector and isovector meson-baryon couplings and the insertion of the fundamental baryon decuplet. }

\,{Likewise, a proposal for future work should contemplate the presence of a crust. 
The density of nuclear matter at the saturation point, corresponding to the minimum value of the nuclear binding energy per nucleon, is estimated by means of nuclear mass analyses, and is of the order of $\rho_0 = 2.8 \times 10^{14}$ g cm$^{-3}$ corresponding to $n_0 = 0.16$ nucleons per fermi cubed. Estimates of the density of the cores of massive neutron stars are in the order of $[5 - 10]\rho_0$. The challenges of consistently describing a neutron star persist in view of the extreme physical conditions of these compact objects, conditions far from those on Earth. In turn, although it is not a crucial point of the present formulation, the outer layer of neutron stars, - the crust -, with density $\rho_{crust} < \rho_0$, presents very different but extremely rich theoretical challenges and observational opportunities insofar as depending on the scenario of its formation, the crust may be very different in its composition and structure. In developing these scenarios for the crust of a neutron star, it is necessary to employ a plethora of theoretical and observational knowledge involving, among others, atomic and plasma physics, the theory of condensed matter, the physics of matter in strong magnetic fields, the theory of nuclear structure, nuclear reactions, the nuclear many-body problem, superfluidity, physical kinetics, hydrodynamics, the physics of liquid crystals, and the theory of elasticity.}

\,{We then pose three fundamental questions involving a noncommutative spacetime:
{\it if} there is a GUP, with a minimum length, {\it then}
(i) does the astrophysical arena represent a relevant laboratory to reveal these effects? (ii) what is the limiting minimum length scale?
(iii) how can we identify GUP effects in observable data from neutron stars?}

\,{Even though we are faced with the limitations of the formulation in view of its preliminary nature,
 we can partially answer these questions: (i) the model predictions are in tune with the scales of the astrophysical arena, which contemplates the range of high energy scales that are beyond terrestrial laboratories and can access unique regimes in compact stars and in cosmology; 
 furthermore, the values of the maximum masses of neutron stars are in tune with the most recent observed values, of the order $[2.0-2.6]M_{\odot}$;
 the results also indicate that, although the effects of the presence of a minimum scale broaden the descriptive perspective of a compact star, encompassing a non-commutative algebra, they do not spoil the effective aspects covered by the QHD formulation; in this sense, the maximum values of pressure and 
 energy density as a function of the baryon density in tune with the QHD-I model predictions stand out;
 (ii) in the present study 
the $\beta$ values varied in a range from $1\times 10^{-5}$ fm$^2$ to
 $5\times 10^{-2}$ fm$^2$; when translated to a minimal length it results in $\Delta x \sim$ 0.003 fm to 0.22 fm. These values can be compared
 with the usual nucleon radius obtained in bag model calculations ranging from 0.2 fm to 1.0 fm \cite{bag1,bag2}, resulting in a limiting minimal length scale at the order of the nucleon's size; (iii) the cutoff values introduced in the EoS has a direct
 inference in limiting the maximum neutron star mass. }
  
 \,{ A more important conclusion to this last question is reached by reasoning in the opposite direction: {\it the fact that neutron stars exist is confirmation that the noncommutative scale must be very small.} Even if it is very small, the minimum length may be relevant in modifying the structure of a primordial phase transition. For example, in the {\it first three minutes of the creation of the universe}~\cite{weinberg1,weinberg2,weinberg3} a cosmological phase transition is believed to have occurred, generating a global change of the primordial matter. Starting at the Planck time $t_P\sim 10^{-44}$ seconds, the young universe evolved and by the time it reached $t\sim 10^{-38}$ seconds the grand unified group $SU(3)\otimes SU (2)\otimes U(1)$ had undergone gauge symmetry breaking. If the minimum length approach is correct, then at this scale the noncommutative effect may play an important role.}

  \,{Additional important aspects to be considered in future formulations of the problem addressed in this contribution concern the thermodynamic consistency of the GUP deformation of spacetime, as well as the obedience to causality, to Le Chatelier's Principle, a fundamental requirement for satisfying equilibrium configurations of a compact star and not spoiling the renormalizability of the original formulation. The QHD-I approach in particular, in its original formulation, is thermodynamically consistent, obeys Le Chatellier's Principle, and is also renormalizable due to the presence of counterterms. The systematic reduction of the phase space due to the presence of the $\beta$ factor raises new questions about the formal consistency of coherent descriptive configurations of a neutron star assuming a noncommutative spacetime, in compliance with these requirements, which deserves this way further studies. }
  
\,{ With particular regard to Le Chatelier's Principle, the matter of the star must satisfy the condition $$dP/d\epsilon \geq 0$$ which is a necessary condition for the stability of a stable star both with respect to its structure as a whole and with respect to the elementary regions of non-equilibrium involving stages of spontaneous contraction or expansion. In our calculations, Le Chatelier's Principle is not completely established, particularly in the tail regions of the pressure curves as a result of the proposed insertion of a minimum length through the GUP deformation, a topic that deserves more attention in the future. Fortunately, the impact of this non-observance of Le Chatelier's Principle does not particularly affect the observation of the effects of a non-commutative algebra on stellar properties, since the adopted model fundamentally contemplates the innermost pressure regions of the star. This limitation serves, however, as motivation for a more in-depth analysis in the future to overcome it.}

\authorcontributions{``Conceptualization, JGGG and DH; methodology, JGGG and DH; software, JGGG and DH; validation, JGGG, DH, POH,MN-M, and CAZV; formal analysis, JGGG, DH, POH,MN-M, and CAZV; investigation, JGGG and DH; resources, JGGG and DH; data curation, JGGG and DH; writing---original draft preparation, DH; writing---review and editing, JGGG, DH, POH,MN-M, and CAZV; visualization, JGGG and DH; supervision, DH; project administration, DH. All authors have read and agreed to the published version of the manuscript.''.}

\acknowledgments{JGGG  was supported by the Coordenação de Aperfeiçoamento de Pessoal de Nível Superior (CAPES) through the Programa de Excelência Acadêmica (Proex). P O H acknowledges financial support from PAPIIT-DGAPA (IN116824). M N-M acknowledges support of the Margarete und Herbert Puschmann-Stiftung.}

\conflictsofinterest{``The authors declare no conflicts of interest.''} 

\appendixtitles{no} 
\appendixstart
\appendix
\section{Expressing Jacobian as a combination of the Poisson brackets}\label{appJac}
In this appendix  we shall summarize the proof of expressing the Jacobian as a combination of the Poisson brackets.
An important canonical invariant is the magnitude of a volume element in
phase space. A canonical transformation $\eta \to \zeta$ transforms the $2N$-dimensional
phase space with coordinates $\eta_i$ to another phase space with coordinates $\zeta_i$ . The
volume element \cite{goldstein}
\bea
(d\eta) = dX_1\, dX_2 \ldots  dX_N\, dP_1\,\ldots\,dP_N
\eea
transforms to a new volume element
\bea
(d\zeta ) = dx_1\, dx_2 \ldots  dx_N\, dp_1\,\ldots\,dp_N\,.
\eea
The sizes of the two volume elements are related by the
absolute value of the Jacobian determinant  $J$
\bea
(d\zeta ) = J\,(d\eta).
\eea
As an example, for $N=1$, the   transformation from    $\eta_i= (X,P)$  to $\zeta_i=(x,p)$ becomes
 \bea
 J&=&\jac{x,p}{X,P }=
 \left|
\begin{array}{cc}
\frac{\pa {x}}{\pa X} & \frac{\pa {x}}{\pa P}  \\
\frac{\pa {p}}{\pa X} & \frac{\pa {p}}{\pa P}  \\
\end{array}
\right|
=\col{x}{p}{X}{P} 
\eea
which results in the volume elements 
\bea
dx\,dp =\col{x}{p}{X}{P}\,\, \,dX\,dP\,.
\eea
For $N=2$ we have the   transformation from   $\eta_i=(X_1,P_1,X_2,P_2)$ to $\zeta_i=(x_1,p_1,x_2,p_2)$
 \bea
 J&=&\jac{x_1,p_1,x_2,p_2}{X_1,P_1,X_2,P_2}=
 \left|
\begin{array}{cccc}
\frac{\pa {x_1}}{\pa X_1} & \frac{\pa {x_1}}{\pa X_2}   &   \frac{\pa {x_1}}{\pa P_1} &  \frac{\pa {x_1}}{\pa P_2}   \\
\frac{\pa {x_2}}{\pa X_1} & \frac{\pa {x_2}}{\pa X_2}   &   \frac{\pa {x_2}}{\pa P_1} &  \frac{\pa {x_2}}{\pa P_2}   \\
\frac{\pa {p_1}}{\pa X_1} & \frac{\pa {p_1}}{\pa X_2}   &   \frac{\pa {p_1}}{\pa P_1} &  \frac{\pa {p_1}}{\pa P_2}   \\
\frac{\pa {p_2}}{\pa X_1} & \frac{\pa {p_2}}{\pa X_2}   &   \frac{\pa {p_2}}{\pa P_1} &  \frac{\pa {p_2}}{\pa P_2}   \\
\end{array}
\right|
\nn\\\nn\\
&=&
\col{x_1}{p_1}{X_1}{P_1} \, \col{x_2}{p_2}{X_2}{P_2}\,
+\col{x_1}{p_2}{X_1}{P_1} \, \col{p_1}{x_2}{X_2}{P_2}\,
\nn\\
&&-\col{x_1}{x_2}{X_1}{P_1} \, \col{p_1}{p_2}{X_2}{P_2}\,
+\col {x_2}{p_2}   {X_1}{P_1} \, \col{x_1}{p_1}  {X_2}{P_2}\,
\nn\\&&
+\col {p_1}{x_2}  {X_1}{P_1} \, \col{x_1}{p_2}   {X_2}{P_2}\,
-\col{p_1}{p_2}    {X_1}{P_1} \, \col  {x_1}{x_2}   {X_2}{P_2}\,\,.
\eea
An extended and  detailed demonstration can be found in the reference of  T. V. Fityo \citep{fityo2008statistical}.
Let us denote $x_i=A_{2i-1}$, $p_i=A_{2i}$, $A_j$ derivative with
respect to $X_i$ we denote $A_{j,2i-1}$, with respect to $P_i$ as
$A_{j,2i}$. Then
\bea
\left\{A_i,A_j\right\}=\sum_{k=1}^D (A_{i,2k-1} A_{j,2k}- A_{i,2k} A_{j,2k-1})\,,
\eea
where the demonstration regards the proof of the following identity
\begin{equation}\label{mnIden}
J=\jac{x_1,p_1,\dots,x_D,p_D}{X_1,P_1,\dots,X_D,P_D} =
\frac1{2^DD!}\sum_{i_1,\dots i_{2D}=1}^{2D} \varepsilon_{i_1\dots
i_{2D}}\{A_{i_1},A_{i_2}\} \dots \{A_{i_{2D-1}},A_{i_{2D}}\},
\end{equation}
where $\varepsilon_{i_1\dots i_{2D}}$ is the Levi-Civita symbol.
\begin{adjustwidth}{-\extralength}{0cm}

\reftitle{References}

\PublishersNote{}
\end{adjustwidth}

\begin{thebibliography}{999}
\bibitem{venez1} Veneziano, G.; A Stringy Nature Needs Just Two Constants. {\em Europhys. Lett.} {\bf 1986}, {\em 2}, 199.
\bibitem{venez2} Amati, D.; Cialfaloni, M.; Veneziano, G.; Superstring collisions at planckian energies.
{\em Phys. Lett. B} {\bf 1987}, {\em 197}, 81.
\bibitem{venez3} Amati, D.; Cialfaloni, M.; Veneziano, G.; Can spacetime be probed below the string size?
{\em Phys. Lett. B} {\bf 1989}, {\em 216}, 41.
\bibitem{gross}  Gross, D. J.;  Mende, P. F. ; The high-energy behavior of string scattering amplitudes. 
{\em Phys. Lett. B} {\bf 1987}, {\em 197}, 129.
\bibitem{lqg1} Rovelli, C.; Loop Quantum Gravity. 
{\em Living Rev. Relativ.} {\bf 2008}, {\em 11}, 5.
\bibitem{lqg2} Ashtekar, A.; New Variables for Classical and Quantum Gravity.
{\em Phys. Rev. Lett.} {\bf 1986}, {\em 57}, 2244.
\bibitem{lqg3}  Rovelli, C.; Smolin, L.; Discreteness of area and volume in quantum gravity.
{\em Nucl. Phys. B} {\bf 1995}, {\em 442}, 593.
\bibitem{qg} Capozziello, S.;  Lambiase, G.; Scarpetta, G.; 
Generalized Uncertainty Principle from Quantum Geometry
{\em Int. Jour. Theor. Phys.} {\bf 2000}, {\em 39}, 15.
\bibitem{heisenberg} Heisenberg, W.; Letter from Heisenberg to Peierls.
{\em  W. Pauli, Scientific Correspondence} {\bf 1995}, {\em Vol. II, Berlin, Springer} 15.
\bibitem{snyder}  Snyder,  H. S.; Quantized Space-Time.
{\em Phys. Rev. } {\bf 1947}, {\em 71}, 38.
\bibitem{szabo}  Szabo, R. J.; Quantum field theory on noncommutative spaces.
{\em Phys. Rep. } {\bf 2003}, {\em 378}, 207.
\bibitem{gamboa}   Gamboa, J.; M\'endez, F.; Loewe,  M.; Rojas,  J. C.;
Noncommutative Quantum Mechanics: The Two-Dimensional Central Field.
{\em Int.  Jour. Mod. Phys. A } {\bf 2002}, {\em 17}, 2555.
\bibitem{wang-yang}Wang, X. J. ; Yan, M. L.; Noncommutative QED and muon anomalous magnetic moment.
 {\em JHEP } {\bf 2002}, {\em 03}, 047.
\bibitem{chaichian} Chaichian, M. ; Sheikh-Jabbari, M. M. ; Tureanu, A.;
Hydrogen Atom Spectrum and the Lamb Shift in Noncommutative QED.
{\em Phys. Rev. Lett. } {\bf 2001}, {\em 86}, 2716.
\bibitem{kempf1}  Kempf, A.; Mangano, G.;  Mann, R. B.; Hilbert space representation of the minimal length uncertainty relation.
{\em Phys. Rev. D } {\bf 1995}, {\em 52}, 1108.
\bibitem{kempf2}  Kempf, A.;  Non-pointlike particles in harmonic oscillators.
{\em J. Phys. A: Math. Gen. } {\bf 1997}, {\em 30}, 2093.
\bibitem{kempf3}  Hinrichsen, H.; Kempf, A.;  Maximal localization in the presence of minimal uncertainties in positions and in momenta.
{\em J. Math. Phys.  } {\bf 1996}, {\em 37}, 2121.
\bibitem{moayedi1}   Moayedi, S. K.; Setare, M. R.;  Moayeri, H.;  
Quantum Gravitational Corrections to the Real Klein-Gordon Field in the Presence of a Minimal Length.
{\em  Int. Jour. Theor. Phys.  } {\bf 2010}, {\em 49}, 2080.
\bibitem{moayedi2}   Moayedi, S. K.; Setare, M. R.;  Moayeri, H.; Poorakbar, M. ; 
Formulation of the Spinor Field in the Presence of a Minimal Length Based on the Quesne Tkachuk Algebra
{\em  Int. Jour. Mod. Phys. A  } {\bf 2011}, {\em 26}, 4981.
\bibitem{dimi1}Ferreira, T. O.; Vasconcellos, C. A. Z. ; Hadjimichef, D. ;
A seesaw-like mechanism for the neutrino in the presence of a minimal length spacetime.
{\em  Astron. Nachr.  } {\bf 2023}, {\em 344}, e220127.
\bibitem{dimi2}Marzola, M. N.; Vasconcellos, C. A. Z. ; Hadjimichef, D. ;
Effects of a generalized uncertainty principle on the MIT bag model equation of state.
{\em  Astron. Nachr.  } {\bf 2024}, {\em 345}, e240016.
\bibitem{Amin} Amin, M.; Mark A. Walton, M.A. ; Quantum-classical dynamical brackets.
{\em Phys. Rev. A} {\bf 2021}, {\em 104}, 032216.
\bibitem{goldstein}  Goldstein, Herbert; Poole, Charles; Safko, John; 
 Classical Mechanics, 3rd Edition,   {\em Ed. Addison-Wesley}, {\bf 2002}.
\bibitem{fityo2008statistical} Fityo, T. V.; Statistical physics in deformed spaces with minimal length.
{\em Phys. Lett. A} {\bf 2008}, {\em 372}, 5872.
\bibitem{qhd1} Walecka, J. D.; A theory of highly condensed matter.
{\em Ann. Phys.} {\bf 1974}, {\em 83}, 491.
\bibitem{qhd2} Serot, B. D.; Walecka, J. D.; The Relativistic Nuclear Many Body Problem.
{\em Adv. Nucl. Phys.} {\bf 1986}, {\em 16}, 1.
\bibitem{qhd3} Matsui, T.; Serot, B. D.; 
The pion propagator in relativistic quantum field theories of the nuclear many-body problem.
{\em Ann. Phys.} {\bf 1982}, {\em 114}, 107.
\bibitem{qhd4} Serot, B. D.; Quantum hadrodynamics.
{\em Rep. Prog. Phys.} {\bf 1992}, {\em 55}, 1855.
\bibitem{qhd5} Serot, B. D.; Walecka, J. D.; Recent Progress in Quantum Hadrodynamics. 
{\em Int. Jour. Mod. Phys. E} {\bf 1997}, {\em 6}, 515.
\bibitem{chang2002effect}Chang, Lay Nam; Minic, Djordje; Okamura, Naotoshi; Takeuchi, Tatsu;
 Effect of the minimal length uncertainty relation on the density of states and the cosmological constant problem.
  {\em Phys. Rev. D}, {\bf 2002}, {\em 65}, 125028. 
\bibitem{rama2001some}Rama, S Kalyana;
Some consequences of the generalised uncertainty principle: statistical mechanical, cosmological, and varying speed of light.
{\em Phys. Lett. B}, {\bf 2001}, {\em 519}, 103. 
\bibitem{bag1} Chodos, A.;  Jaffe, R. L.;  Johnson, K.;  Thorn, C. B.;  Weisskopf, V. F.;
 New extended model of hadrons.
{\em Phys. Rev. D}, {\bf 1974}, {\em 9}, 3471. 
\bibitem{bag2}Goldflam, R. ; Wilets, L.; Soliton bag model.
{\em Phys. Rev. D}, {\bf 1982}, {\em 25}, 1951. 
\bibitem{weinberg1} Weinberg, S.;
The First Three Minutes: A Modern View of the Origin of the Universe 
{\em Ed.	Basic Books}, {\bf 1993}.
\bibitem{weinberg2} Weinberg, S.; Gravitation and Cosmology: Principles and Applications of the General 
Theory of Relativity.  {\em Ed. John Wiley \& Sons}, {\bf 1972}.
\bibitem{weinberg3} Weinberg, S.; Cosmology.
 {\em Ed. Oxford University Press}, {\bf 2008}.
 \bibitem{Mandl} Franz Mandl and Graham Shaw;
Quantum Field Theory.
    {\em Wiley}, {\bf 2010}.
\end{thebibliography}
\end{document}